\documentclass[conference]{IEEEtran}
\IEEEoverridecommandlockouts
\usepackage{cite}
\usepackage{amsmath,amssymb,amsfonts}
\usepackage{algorithmic}
\usepackage{graphicx}
\usepackage{textcomp}
\usepackage{xcolor}
\usepackage{comment}

\usepackage{fancyhdr}

\usepackage{caption}
\usepackage{siunitx}
\usepackage{booktabs}
\usepackage{tabularx}
\newcolumntype{Y}{>{\centering\arraybackslash}X}

\def\BibTeX{{\rm B\kern-.05em{\sc i\kern-.025em b}\kern-.08em
    T\kern-.1667em\lower.7ex\hbox{E}\kern-.125emX}}
\begin{document}

\bstctlcite{IEEEexample:BSTcontrol}

\title{The First IEEE UV2022 Mathematical Modelling Competition: Backgrounds and Problems}
\makeatletter 
\newcommand{\linebreakand}{%
  \end{@IEEEauthorhalign}
  \hfill\mbox{}\par
  \mbox{}\hfill\begin{@IEEEauthorhalign}
}
\makeatother 

\author{
\IEEEauthorblockN{Juntao Jiang}
\IEEEauthorblockA{\textit{Universal Village Society} \\
Boston, USA \\
jtjiang@universal-village.org}

\and
\IEEEauthorblockN{Yuan Niu}
\IEEEauthorblockA{\textit{Universal Village Society} \\
Boston, USA \\
niuyuan@citics.com}

\and
\IEEEauthorblockN{Yi Tao}
\IEEEauthorblockA{\textit{Universal Village Society} \\
Boston, USA \\
yitao@universal-village.org}
}

\maketitle

\thispagestyle{fancy}
\fancyhead{} 
\fancyfoot{}
\rhead{}
\cfoot{} 
\rfoot{}

\begin{abstract}
Economic growth, people's health, and urban development face challenges in the post-epidemic era. How to promote high-quality and sustainable urban development, improve citizens' sense of  happiness, and solve problems in city management have become a heated and crucial topic. Mathematical modeling is a research method that uses symbols to express practical concerns, establish mathematical models, and propose solutions. The 1$^{st}$ IEEE UV2022 Mathematical Modelling Competition is a satellite activity of the 6$^{th}$ IEEE International Conference on Universal Village, which expects participants to use mathematical modeling methods for practical problems and provide guidelines for sustainable social progress. This short paper introduces the background of the competition and publishes the problems to be solved.
\end{abstract}

\hfill  


\section{Introduction}

Our current world is more complex than ever: emerging technologies like artificial intelligence, virtual reality, artificial reality, big data, cloud computing and 5G continue to be developed, changing people’s daily life and providing a solid foundation for more substantial growth; the severe pandemic has made an impact on politics \cite{rapeli2020will,landman2020pandemic}, economics \cite{maital2020global}, education \cite{setiawan2020covid}, lifestyles \cite{hashem2020examining} and resident adaptations \cite{chertoff2020early}; structural vulnerabilities and dynamic inequalities have been significantly enhanced \cite{leach2021post}; the post-pandemic era economic recovery has not yet begun, but health risks still exist; the energy crisis continues to affect supply chain security \cite{hutter2022russia}. These challenges are what all countries need to face. At the same time, different countries face different opportunities and problems due to their different cultural traditions, development processes, and geopolitics. We hope to express these complex situations simply and abstractly. In general, the difficulties faced by developing countries and economically less developed regions are more severe than those developed ones. We hope to express these complex situations simply and abstractly. We hope to provide inspiring ideas from macro policies to specific issues. We hope to see effective, practical, comprehensive, and innovative solutions.

Universal Village is a new concept proposed by MIT’s Universal Village Program, which advocates promoting harmony between man and nature through the prudent use of technology and addressing the environmental challenges brought about by rapid urbanization \cite{cao2018preliminary}. It is also the original intention of the UV conference series to comprehensively use various technologies to achieve the goal of an ideal society. The 6$^{th}$ IEEE International Conference on Universal Village (UV2022) features the theme of “Post-Pandemic Reflection on Health, Harmony, and Sustainability: Mobility and Virtual Connection; Diversity and System Efficiency; Responsiveness and Resilience; Inclusiveness and Integration,” focuses on essential topics in the post-pandemic era.

As a satellite activity of IEEE UV2022, the 1$^{st}$ IEEE UV2022 Mathematical Modelling Competition is held to use mathematical modeling methods to solve practical problems. The availability of fast and powerful computers has made it possible to mathematize complex problems in industry and commerce\cite{towers2020guide}  and solve them better. Common mathematical modeling problems include optimization, evaluation, prediction, etc.; the commonly used methods include integer programming, linear programming, nonlinear programming, graph theory, analytic hierarchy process, regression prediction, principal component analysis, etc.

This short paper officially publishes the problems of the 1$^{st}$ IEEE UV2022 Mathematical Modelling Competition. Participants should choose one problem, carefully analyze the competition problems, understand the relevant background, search and organize related material, build mathematical models, write programs to solve the models, and complete report writing. The paper needs to contain the abstract, the introduction/background, the problem statement, details of models and algorithms, the sensitivity analysis, strengths and weaknesses, and the conclusion.

\section{Problem A: Smart City Development Index}
\subsection{Background}
The world population living in urban areas will increase to $66\%$ by 2030, according to UN\cite{un}. The level of city development is directly related to the quality of human life. The Smart City is an evolving concept about improving the function of cities using information and communication technologies \cite{batty2012smart}. As the increasing population, pollution, congestion, resource usage, and increasingly stricter energy and environmental requirements continue to affect life qualities \cite{chourabi2012understanding}, smart cities nowadays should be able to apply new technologies to solve or alleviate these problems.

A fair, reasonable, and comprehensive city development evaluation index can help compare different cities' situations and guide today's urban construction.

Take Hangzhou as an example. In 2016, Hangzhou created the first "city brain" in China. Driven by this, the pace of Hangzhou's exploration of urban digital construction has been accelerating. At the city-wide digital economy high-quality development conference held in September 2022, Hangzhou proposed to build the city with the highest digital economy development level in China. Similarly, facing the difficulties of urban operation and management, Harbin is constantly deepening and expanding smart application scenarios, realizing smart governance of the city through innovation and deepening the construction of smart applications.

The evaluation should be done in the following aspects, which are known as UV subsystems.
\begin{itemize}
  \item {Smart Home and Community}
  \item {Smart Medicine and Healthcare \cite{zhang2020evaluation}}
   \item {ITS, Urban Planning and Crowd Management \cite{xu2020evaluation}}
   \item {Smart Energy Management \cite{yang2020evaluation}}
   \item {Smart City Infrastructure \cite{wu2020evaluation}}
   \item {Smart Response System for City Emergency \cite{yang2020evaluation}}
   \item {Smart Environmental Protection \cite{yuan2020evaluation}}
   \item {Smart Humanity \cite{cao2020evaluation}}
\end{itemize}

Problem A focuses on building an index for smart city development evaluation and applying this index to Hangzhou and Harbin.

\subsection{Tasks}
\subsubsection{Task 1}

Define a "Smart City Development Index" as a metric to measure the success of smart city development. We encourage the participant to consider all eight UV subsystems in the index.

\subsubsection{Task 2}

Research the recent development of Hangzhou and Harbin. Use the proposed metric to evaluate the development level for these two cities.

\subsubsection{Task 3} 

Choose a city in a country other than China and research the recent development. Use the proposed metric to evaluate the development level of this city.

\subsubsection{Task 4}

Predict the future change in each subsystem in the next ten years. Predict the future change in Hangzhou and Harbin's proposed "Smart City Development Index" value in the next ten years. 

\subsubsection{Task 5}

Based on situations in Hangzhou and Harbin, make development proposals and formulate plans for these two cities.

\subsection{Possible Useful Links}
1) \textit{Hangzhou Statistical Yearbook:}

\emph{http://tjj.hangzhou.gov.cn/col/col1229453592/index.html} \\

2) \textit{Harbin Statistical Yearbook:} 

\emph{http://www.harbin.gov.cn/col/col39/index.html}
\section{Problem B: Vaccine Allocation}
\subsection{Background}
The pandemic in the past three years has brought huge disasters to human beings and changed people's way of life. The emergence of the Omicron variant of SARS-CoV-2 last winter made the epidemic spread more quickly. Vaccines, which have saved tens of millions of lives globally \cite{watson2022global}, remain the most important method for controlling COVID-19 and shifting the pandemic to the next phase \cite{del2022winter}. 

With the change in China's pandemic control policies, it is essential to promote vaccination, especially among the elderly.
We expect to open more vaccination points in central hospitals, community hospitals, and health centers to facilitate vaccinating citizens. However, due to the cost of vaccine transportation and storage, we must consider how to distribute vaccines to central hospitals, community hospitals, and health centers.

Problem B focuses on designing a reasonable vaccine allocation plan to ensure the vaccination demand and consider the cost issue.

\subsection{Tasks}
\subsubsection{Task 1}

Predict and visualize national daily vaccination numbers for the next three months.

\subsubsection{Task 2}

Considering the number of nearby residents, transportation convenience, number of medical staff, vaccine storage and transportation costs, and  avoiding excessive gathering of people during vaccination, design a vaccine allocation plan for central hospitals, community hospitals, and health centers. 

\subsubsection{Task 3}

Taking Hangzhou Gongshu District and Harbin Daoli District as examples, calculate the number or proportion of vaccines distributed to each central hospital, community hospital, and health center in the two districts. 

\subsubsection{Task 4}

Briefly write a note on vaccine allocation (e.g., prioritizing the elderly.)

\subsection{Possible Useful Links}
1) National COVID-19 vaccination status:

\emph{http://www.nhc.gov.cn/xcs/yqjzqk/list\_gzbd.shtml}

\section{Problem C: LinkNYC in China}
\subsection{Background}
In 2016, the New York City government and Google-backed CityBridge jointly launched and built a public communication project - LinkNYC, to redesign telecommunication to activate the "Twenty-First-Century Creative City" \cite{maier20183}. It appears as kiosks on New York streets where people can get free Wi-Fi, charge their phones, use city services and maps for directions, and make free calls within the U.S. and to emergencies. It is 10 feet tall and is equipped with displays, cameras, tablets, speakers, microphones, and sensors. The original intention of creating LinkNYC was to make the city better meet the needs of citizens. 

LinkNYC's kiosk is shown in Figure 1. The functions include: 
1) using personal devices to connect to LinkNYC’s super free Wi-Fi.
2) getting access to city services, maps, and directions from the tablet; 3) making free phone calls anywhere in the U.S, using the tablet or the tactile keypad and microphone, plugging in personal headphones for more privacy; 4)
using the dedicated red 911 button in the event of an emergency; 5) charging your device in a power-only USB port; 6) enjoying more room on the sidewalk with Link’s sleek, ADA-compliant design; 7) viewing public service announcements and more relevant advertising on two 55” HD displays

\begin{figure}
	\begin{minipage}{0.8\linewidth}
		\vspace{3pt}
		\centerline{\includegraphics[width=\textwidth]{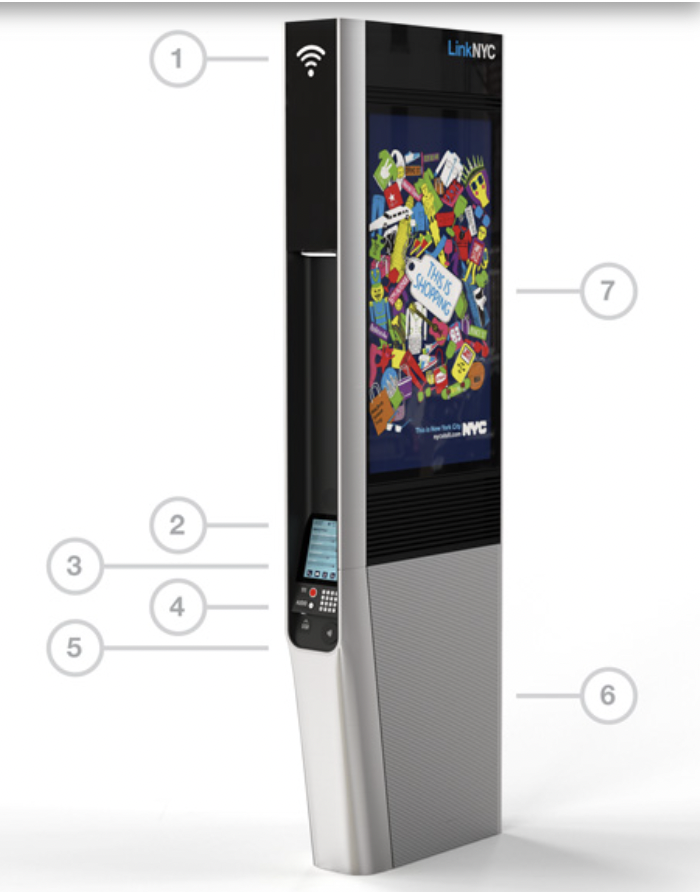}}
	\end{minipage}
\caption{LinkNYC's kiosks, the figure is from \cite{linknyc}. }
\end{figure}

Problem C focuses on estimating how many such kiosks are needed in a city. Also, it is expected to design a sustainable profit model.

\subsection{Tasks}
\subsubsection{Task 1} 

If we introduce the LinkNYC to China and build information kiosks in Hangzhou and Harbin, approximately how many information kiosks need to be made to meet the needs of citizens and avoid the waste of resources? Please estimate the number of kiosks required in each district of Hangzhou and Harbin.

\subsubsection{Task 2}

Please design the functions included in the kiosk introduced in China. We hope that the information kiosk includes as many free convenience functions as possible and brings profits through commercial models such as advertisements and some paid functions. Please create a profit model to illustrate.

\subsubsection{Task 3}

Please specify the upper time limit for each user to use the kiosk so that everyone can fully use the service and avoid others waiting for a long time.

\subsubsection{Task4}

Please give your suggestions on promoting this information kiosk in Chinese cities.

\subsection{Possible Useful Links}
1) Official Site of LinkNYC: \emph{https://www.link.nyc/}

\section{Conclusion}
This paper introduces the background and problems of  the 1$^{st}$ IEEE UV2022 Mathematical Modelling Competition, a satellite activity of the 6$^{th}$ IEEE International Conference on Universal Village. The competition aims to call for solutions based on mathematical modeling methods for real-world problems. The problems are the smart city development index design, the vaccine allocation problem, and the introduction of LinkNYC kiosks to China. Participants are expected to choose one problem, according to the background and each task, then do the analysis, modeling, programming, and writing.
\bibliographystyle{IEEEtran}
\bibliography{final}

\begin{thebibliography}{10}
\providecommand{\doi}[1]{doi:#1}
\providecommand{\webdate}[1]{accessed #1}
\providecommand{\url}[1]{#1}
\csname url@samestyle\endcsname
\providecommand{\newblock}{\relax}
\providecommand{\bibinfo}[2]{#2}
\providecommand{\BIBentrySTDinterwordspacing}{\spaceskip=0pt\relax}
\providecommand{\BIBentryALTinterwordstretchfactor}{4}
\providecommand{\BIBentryALTinterwordspacing}{\spaceskip=\fontdimen2\font plus
\BIBentryALTinterwordstretchfactor\fontdimen3\font minus
  \fontdimen4\font\relax}
\providecommand{\BIBforeignlanguage}[2]{{%
\expandafter\ifx\csname l@#1\endcsname\relax
\typeout{** WARNING: IEEEtran.bst: No hyphenation pattern has been}%
\typeout{** loaded for the language `#1'. Using the pattern for}%
\typeout{** the default language instead.}%
\else
\language=\csname l@#1\endcsname
\fi
#2}}
\providecommand{\BIBdecl}{\relax}
\BIBdecl

\bibitem{rapeli2020will}
L.~Rapeli and I.~Saikkonen, ``How will the covid-19 pandemic affect
  democracy?'' \emph{Democratic Theory}, vol.~7, no.~2, pp. 25--32, 2020.

\bibitem{landman2020pandemic}
T.~Landman and L.~D.~G. Splendore, ``Pandemic democracy: elections and
  covid-19,'' \emph{Journal of Risk Research}, vol.~23, no. 7-8, pp.
  1060--1066, 2020.

\bibitem{maital2020global}
S.~Maital and E.~Barzani, ``The global economic impact of covid-19: A summary
  of research,'' \emph{Samuel Neaman Institute for National Policy Research},
  vol. 2020, pp. 1--12, 2020.

\bibitem{setiawan2020covid}
B.~Setiawan and V.~Iasha, ``Covid-19 pandemic: The influence of full-online
  learning for elementary school in rural areas,'' \emph{JPsd (Jurnal
  Pendidikan Sekolah Dasar)}, vol.~6, no.~2, pp. 114--123, 2020.

\bibitem{hashem2020examining}
T.~N. Hashem, ``Examining the influence of covid 19 pandemic in changing
  customers' orientation towards e-shopping,'' \emph{Modern Applied Science},
  vol.~14, no.~8, pp. 59--76, 2020.

\bibitem{chertoff2020early}
J.~D. Chertoff \emph{et~al.}, ``The early influence and effects of the
  coronavirus disease 2019 (covid-19) pandemic on resident education and
  adaptations,'' \emph{Journal of the American College of Radiology}, vol.~17,
  no.~10, pp. 1322--1328, 2020.

\bibitem{leach2021post}
M.~Leach, H.~MacGregor, I.~Scoones, and A.~Wilkinson, ``Post-pandemic
  transformations: How and why covid-19 requires us to rethink development,''
  \emph{World Development}, vol. 138, p. 105233, 2021.

\bibitem{hutter2022russia}
C.~Hutter and E.~Weber, ``Russia-ukraine war: Short-run production and labour
  market effects of the energy crisis,'' IAB-Discussion Paper, Tech. Rep.,
  2022.

\bibitem{cao2018preliminary}
S.~Cao \emph{et~al.}, ``Preliminary study on evaluation of smart-cities
  technologies and proposed uv lifestyles,'' in \emph{2018 4th International
  Conference on Universal Village (UV)}.\hskip 1em plus 0.5em minus 0.4em\relax
  IEEE, 2018, pp. 1--49.

\bibitem{towers2020guide}
D.~A. Towers, D.~Edwards, and M.~Hamson, \emph{Guide to mathematical
  modelling}.\hskip 1em plus 0.5em minus 0.4em\relax Bloomsbury Publishing,
  2020.

\bibitem{un}
``The world's cities in 2016,'' \url{http://www.un.org/ (2016)}.

\bibitem{batty2012smart}
M.~Batty \emph{et~al.}, ``Smart cities of the future,'' \emph{The European
  Physical Journal Special Topics}, vol. 214, no.~1, pp. 481--518, 2012.

\bibitem{chourabi2012understanding}
H.~Chourabi \emph{et~al.}, ``Understanding smart cities: An integrative
  framework,'' in \emph{2012 45th Hawaii international conference on system
  sciences}.\hskip 1em plus 0.5em minus 0.4em\relax IEEE, 2012, pp. 2289--2297.

\bibitem{zhang2020evaluation}
L.~Zhang \emph{et~al.}, ``Evaluation of smart healthcare systems and novel
  uv-oriented solution for integration, resilience, inclusiveness and
  sustainability,'' in \emph{2020 5th International Conference on Universal
  Village (UV)}.\hskip 1em plus 0.5em minus 0.4em\relax IEEE, 2020, pp. 1--28.

\bibitem{xu2020evaluation}
L.~Xu \emph{et~al.}, ``Evaluation of transportation systems and novel
  uv-oriented solution for integration, resilience, inclusiveness and
  sustainability,'' in \emph{2020 5th International Conference on Universal
  Village (UV)}.\hskip 1em plus 0.5em minus 0.4em\relax IEEE, 2020, pp. 1--63.

\bibitem{yang2020evaluation}
Z.~Yang \emph{et~al.}, ``Evaluation of smart energy management systems and
  novel uv-oriented solution for integration, resilience, inclusiveness and
  sustainability,'' in \emph{2020 5th International Conference on Universal
  Village (UV)}.\hskip 1em plus 0.5em minus 0.4em\relax IEEE, 2020, pp. 1--49.

\bibitem{wu2020evaluation}
S.~Wu \emph{et~al.}, ``Evaluation of smart infrastructure systems and novel
  uv-oriented solution for integration, resilience, inclusiveness, and
  sustainability,'' in \emph{2020 5th International Conference on Universal
  Village (UV)}.\hskip 1em plus 0.5em minus 0.4em\relax IEEE, 2020, pp. 1--45.

\bibitem{yuan2020evaluation}
H.~Yuan \emph{et~al.}, ``Evaluation of smart environmental protection systems
  and novel uv-oriented solution for integration, resilience, inclusiveness and
  sustainability,'' in \emph{2020 5th International Conference on Universal
  Village (UV)}.\hskip 1em plus 0.5em minus 0.4em\relax IEEE, 2020, pp. 1--77.

\bibitem{cao2020evaluation}
S.~Cao \emph{et~al.}, ``Evaluation of smart humanity systems and novel
  uv-oriented solution for integration, resilience, inclusiveness and
  sustainability,'' in \emph{2020 5th International Conference on Universal
  Village (UV)}.\hskip 1em plus 0.5em minus 0.4em\relax IEEE, 2020, pp. 1--28.

\bibitem{watson2022global}
O.~J. Watson \emph{et~al.}, ``Global impact of the first year of covid-19
  vaccination: a mathematical modelling study,'' \emph{The Lancet Infectious
  Diseases}, vol.~22, no.~9, pp. 1293--1302, 2022.

\bibitem{del2022winter}
C.~Del~Rio, S.~B. Omer, and P.~N. Malani, ``Winter of omicron—the evolving
  covid-19 pandemic,'' \emph{Jama}, vol. 327, no.~4, pp. 319--320, 2022.

\bibitem{maier20183}
M.~Maier, M.~Mcbride, and P.~Mcconnell, ``3. linknyc: Redesigning
  telecommunication to activate the twenty-first-century creative city,'' in
  \emph{Smarter New York City}.\hskip 1em plus 0.5em minus 0.4em\relax Columbia
  University Press, 2018, pp. 79--106.

\bibitem{linknyc}
``Linknyc,'' \url{https://www.link.nyc/}.

\end{thebibliography}

\newpage

\vspace{12pt}
\color{red}

\end{document}